\documentclass[%
 reprint,
 amsmath,
 amssymb,
 prstab,
]{revtex4-1}

\usepackage{graphicx}
\usepackage{dcolumn}
\usepackage{bm}
\usepackage{subfigure}
\usepackage{bm}
\usepackage{array,color}
\usepackage{multirow}
\usepackage{amsmath}
\usepackage[none]{hyphenat}

\usepackage[mathlines]{lineno}

\begin{document}

\preprint{APS/123-QED}

\title{A new description of space charge induced beam halo} 

\author{Chao Li}
\altaffiliation[lichao@ihep.ac.cn]{}
\affiliation{%
Key Laboratory of Particle Acceleration Physics and Technology, Institute of High Energy Physics, Chinese Academy of Sciences, 19(B) Yuquan Road, Beijing 100049, China}
\affiliation{%
Institut f\"{u}r Kernphysik--4, Forschungszentrum J\"{u}lich, 52425 J\"{u}lich, Germany}


\author{R. A. Jameson}%
\affiliation{%
Institut f\"{u}r Angewandte Physik, Goethe Uni Frankfurt, Frankfurt-am-Main, Germany}




\begin{abstract}

In particle accelerators, particle losses to surrounding components must be minimized.  These particles usually have been driven into an undesired ``halo" outside the desired distribution before being lost. The processes involved and physical identification of such halo is an intense research topic especially when the nonlinear space charge effect is taken into account.  In order to present a method,  firmly based on the physics of the transient beam state,  for determining when a particle may be defined as being in a halo, the particular example of an initially rms matched beam injected into a resonance of an alternating-gradient (AG) focusing lattice is studied, in which the resonance between the collective mode and single particle motion pushes some particles into a halo defined by extending the halo description to the action-angle frame. In this new  frame of description, the halo particles can be physically defined as those with larger actions affected by the outer quasi hyperbolic tori, which contains the halo particles observed in the traditional approaches on projected 2D  phase space, as well as some particles still hidden in the core.



\begin{description}
\item[PACS numbers] 41.75.-i, 29.27.Bd, 29.20.Ej \\
\end{description}
\end{abstract}


\maketitle

In megawatt class high intensity ion accelerators, small particle losses from the formation of beam halo or tail particles contribute to induced component radioactivity and must be strictly minimized.  The basic physical understanding of the beam halo formation is crucial and provides guidelines for halo observation and possible prevention \cite{10}. 
Although researches are aware of beam halo particles and can identify them in experiments \cite{aleksandrov2016beam}, there is still no consensus for a clear and universal description of halo relevant for any beam distribution type academically. As to ways of halo description, typical example is that the halo particles are depicted by those are ``far from" or ``close to" the beam ``core"-- subjective geometrical   characteristics. Researchers also explored other approaches to depict the severity of the beam halo in numerical simulation. Examples are ``halo parameter" defined as function of the fourth order moments and the second order moments of the particle distribution \cite{4}; ``core-halo-limit" based on the location where the density gradient in real space abruptly changes from small variations in the halo to a very steep variation when arriving on the ``wall" of the core uniform distribution \cite{5}. However, these methods starting from different assumptions only indicate partial characteristics of  halo formation.

Generally, the beam halo generation are considered as consequence of nonlinear effect from external field and self-induced space charge \cite{chen1991degradation, 3, ekdahl2017emittance}. In this paper, heam halo due to space charge will be discussed in detail since it requires self-consistent treatment. The observation of space charge induced beam halo is mostly attributed to certain ``resonance" effects. Simplifications such as smoothed channel and linear space charge assumption inside the beam core were adopted to model the halo particles in the past and most of these models were facing some basic difficulties, even when they were only used to explain observation obtained from numerical simulations. As an extension of the former research \cite{6,23}, in this paper, with coasting beam assumption in an non-smoothed AG focusing channel, we discuss a new description of a halo boundary and halo particles based on the physical properties of the action-angle phase space.

As to ``resonance" phenomena in accelerators, the motions of single particle $x(s)$, beam envelope oscillation $R(s)$ and collective modes $I_{j;k,l}(s)$ have to be studied simultaneously, among which several resonances can be excited. Inheriting the terminologies previously defined \cite{6, 23}, $\Phi_e$ and $\Phi_{j;k,l}$ represent the phase advance of matched beam envelope oscillation ($360^{\circ}$ per period) and the collective motion $I_{j;k,l}$ respectively; $\sigma_{0}$ and $\sigma$ are the standard phase advances, representing the average focusing strengths particles feel without and with space charge; $\sigma_s$ is the single particle phase advance, which is usually clustered around $\sigma$ \footnote{The nonlinear effects from external elements and internal space charge cause different particles to have different particle phase advance $\sigma_s$, leading to a beam $phase$ $advance$ $spread$. In rings, this phase advance spread in equivalent to $tune$ $spread$.}. All of these phase advances are relative to (per) the same focusing period length for consistency with $\Phi_e$.

\begin{figure}
    \centering
    \subfigure[]{
        \label{fig:subfig:4.1.1} 
        \includegraphics[width=3in]{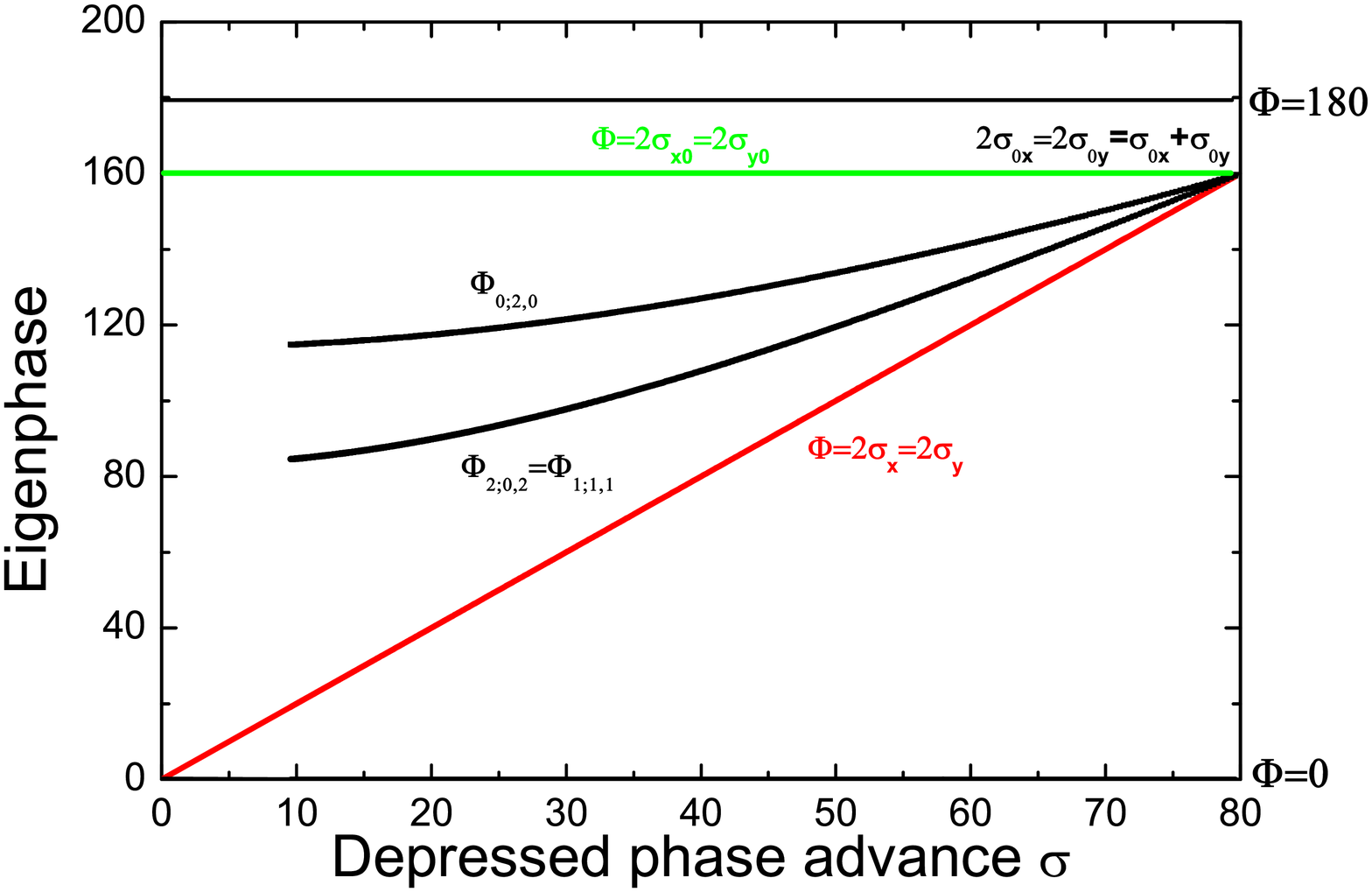}}
    \subfigure[ ]{
        \label{fig:subfig:4.1.2} 
        \includegraphics[width=3in]{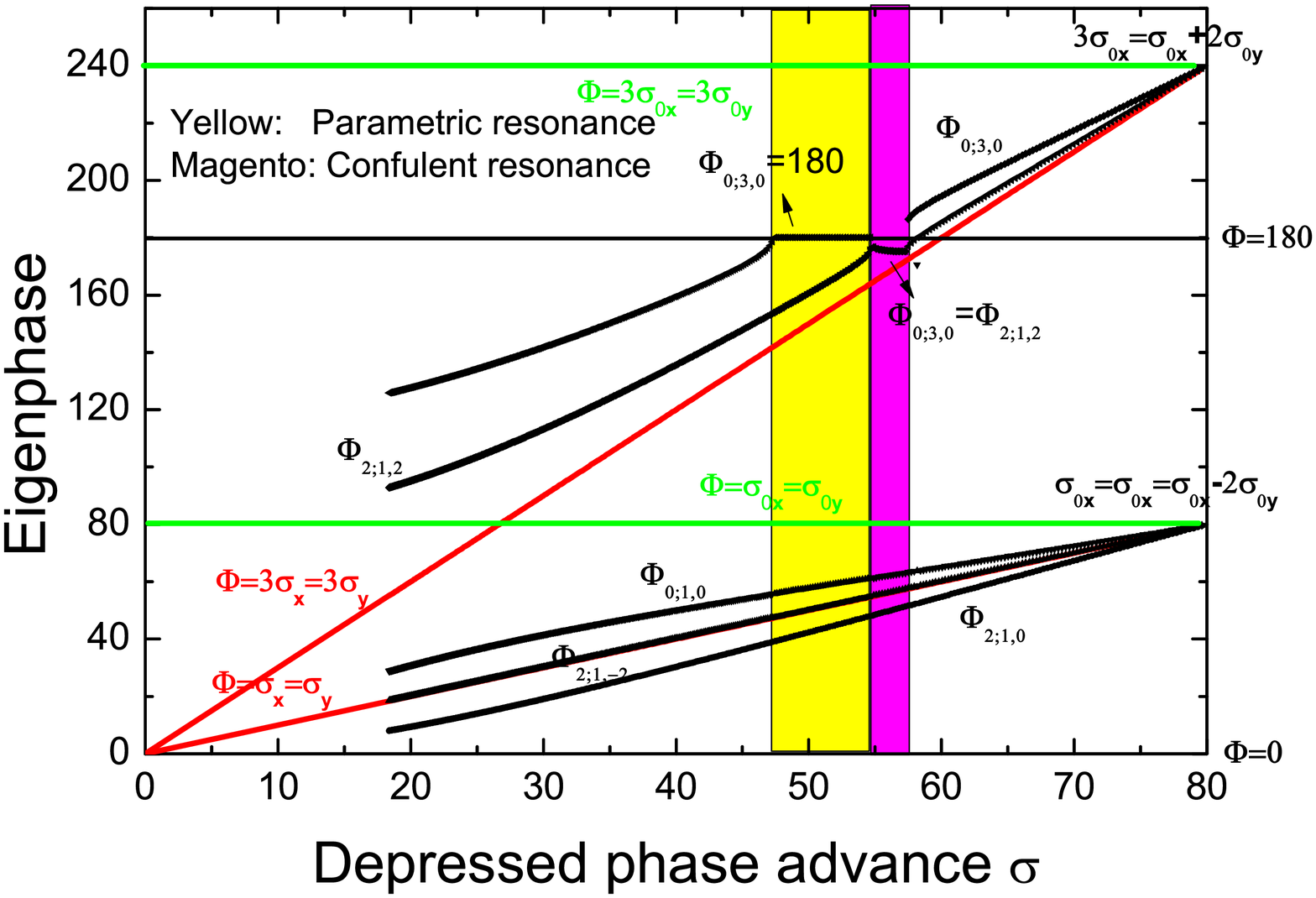}}
    \subfigure[]{
        \label{fig:subfig:4.1.3} 
        \includegraphics[width=3in]{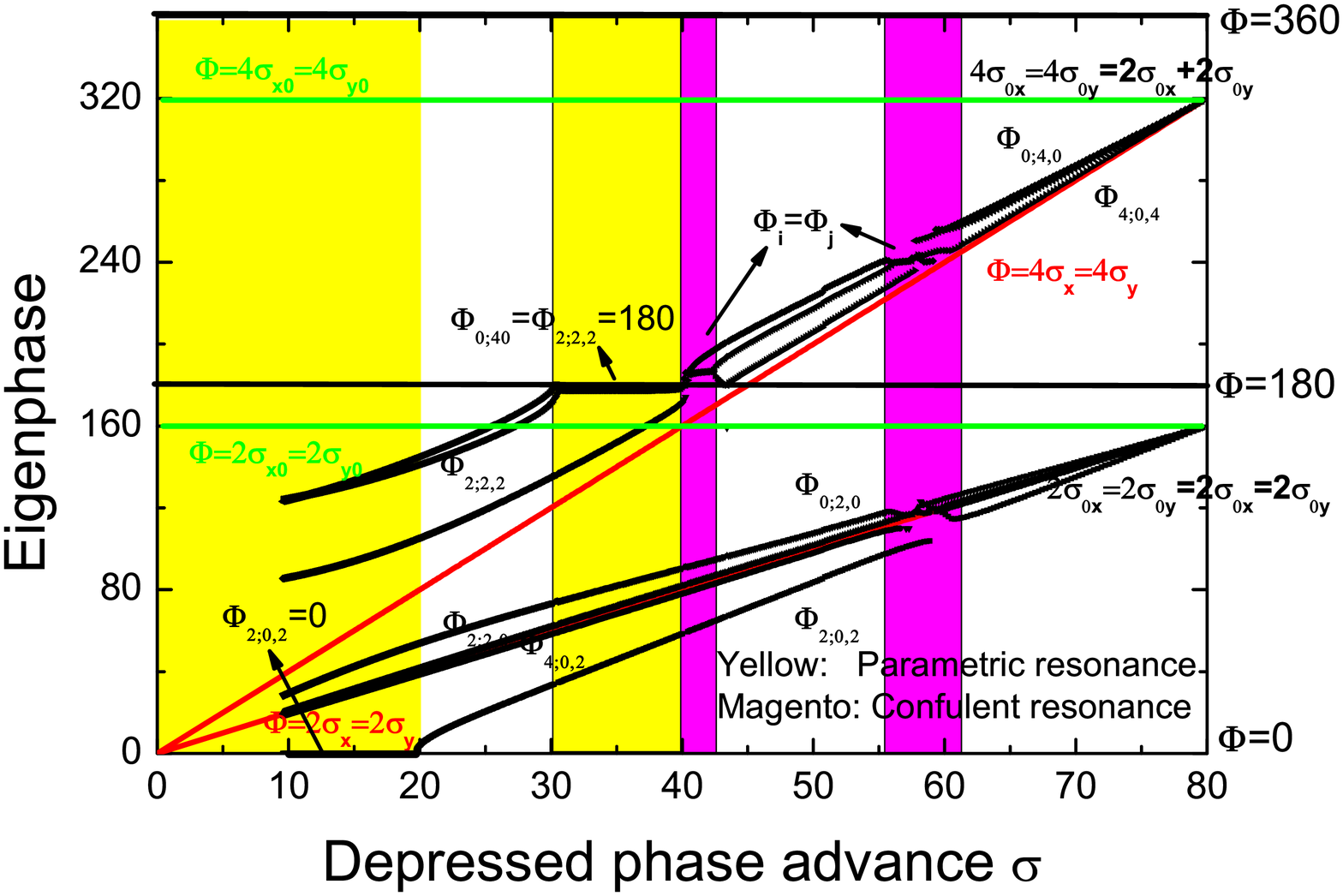}}
        \caption{\label{fig:4.1}  Eigenphases of the 2nd (a), 3rd (b), 4th (c) orders collective modes in symmetric FODO channel, with $\epsilon_x=\epsilon_y$ and $\sigma_{x0}=\sigma_{y0}=\sigma_0=80^{\circ}$. Coloured area is collective stop band.} 
\end{figure}

For an initially rms matched beam injected into an AG focusing channel, one of the resonances is the particle-core resonance \cite{10,15,16,17}, $\sigma_s/\Phi_e=m/n$, which has been regarded as one of the potential mechanisms for beam halo formation. Previous work based on a simple particle-core model shows the interactive processes between the matched oscillating ``core" and halo particles; however, this particle-core model lacks self-consistence  and gives only limited  predictions. It has been shown that this ``particle-core resonance" will cause beam halo, but over a long time scale \cite{12}. Another resonance is the structure resonance, which is well-explained in the former work \cite{6,23,9}. The structure resonance is made up of $confluent$ $resonance$ -- taking place between different $\Phi_{j;k,l}$, and $parametric$ $resonance$ when $\Phi_{j;k,l}/\Phi_e=m/n$. Fig.~\ref{fig:4.1} shows the eigenphases of the 2nd, 3rd, 4th orders of collective eigenphases $\Phi_{j;k,l}$ as function of depressed phase advance $\sigma$, with $\sigma_0=80^\circ$. The coloured areas where the structure resonances take place are termed collective instability stop bands.

The third set of resonance takes place between single particle and collective mode $\sigma_s/\Phi_{j;k,l}=m/n$ -- termed  particle--collective mode resonance. In this paper, this mechanism is used to depict the halo particle generation. In fact, the interaction between particle and 2nd order even collective modes  ($\Phi_{2;0,2}$ and $\Phi_{0;2,0}$) \cite{12,13,14} has been preliminarily  discussed so far only in the constant focusing channel.  It has been verified that if the particles were only affected by only one mode with specific condition \footnote{By limiting the beam distribution type to keep beam only being affected by the lower mode $\phi_{2;0,2}$, or by  using an appropriately anti-symmetric beam condition $\epsilon_x \not\equiv \epsilon_y$ to move the mode $\Phi_{2;0,2}$ out of the region between red line and green line in Fig.~\ref{fig:subfig:4.1.1}. One example of the latter method is also described as the fixed point of the quadrupole mode has dropped into the core \cite{20}.},  the manifold structure at a Poincar$\acute{e}$ surface plot near the beam surface will be modified. The possibility for a single particle being captured by mixed resonances ($\sigma_s/\Phi_{2;0,2}=m/n$, or $\sigma_s/\Phi_{0;2,0}=m/n$), or being lost in a chaotic region due to resonances overlapping can be reduced \cite{18,19,20}. In the following, we extend the discussion on resonance between particle and collective modes to arbitrary order and the halo particles are described in a new frame: action-angle ($J,\phi$) phase space.

Express the perturbed space charge potential leading to the $nth$ order collective mode as 
\begin{eqnarray}\label{eq1}
V_n(x,y;s)&=&\sum_{j=0}^{n}A_j(s)x^{n-j}y^j + \sum_{j=0}^{n-2}...,
\end{eqnarray}
which is  exactly the same as pseudomultipoles in rings except that the coefficients $A_j(s)$ are time dependent. This indicates that the perturbed space charge potential has to be self-consistently evolving with the beam. If only the leading terms (first term in the right side in Eq.~(1)) are kept, the generalized single particle Hamiltonian is
\begin{eqnarray}\label{eq2}
& & H_p(x,p_x,y,p_y;s) = \nonumber \\ 
&=& \frac{1}{2}(K_x(s)x^2+p_x^2)+ \frac{1}{2}(K_y(s)y^2+p_y^2)+V_n(s).
\end{eqnarray}
where $K_{x,y}(s)$ represent the external focuisng strength. 
In Courant-Snyder description, new pairs of canonical conjugate variables can be obtained with the transformation $x=\sqrt{\beta_x J_x} \cos \Phi_x$, where $\Phi_x=\phi_x+\chi_x-\nu_x \theta$ with $\chi_x=\int_0^s1/\beta_x ds$, $\nu_x=\sigma_x/360^{\circ}$, $\theta=2\pi s /L$; similar transformation on $y$ direction. The Hamiltonian turns into
\begin{eqnarray}\label{eq3}
&&J(\Phi_x,J_x, \Phi_y,J_y;\theta) = \nu_x J_x +\nu_y J_y +\sum_{j=1}^n \sum_{k,l}V_{k,l}(\theta) 
\end{eqnarray}
Here, $V_{k,l}(\theta)=A_j(\theta)\bar{V}_{k,l}(J_x \beta_x(\theta),J_y \beta_y(\theta))\cos(k\Phi_x+l\Phi_y)$ is a periodic function of $\theta$ and can be expanded in Fourier harmonics
\begin{eqnarray}\label{eq4}
V_n&=&\sum_{j=1}^n\sum_{k,l}V_{k,l}(\theta)\nonumber \\
&=&\sum_{j=1}^n\sum_{k,l}\sum_{p}G_{k,l;p}e^{i(k\phi_x+l\phi_y-p\theta+k\chi_x-l\chi_y)},
\end{eqnarray}
where
\begin{eqnarray}\label{eq5}
G_{k,l;p}=\frac{1}{2\pi}\int_0^L d\theta A_j(\theta)\bar{V}_{k,l}e^{-i(k\nu_x+l\nu_y-p)\theta}
\end{eqnarray}
is the Fourier amplitude which represents the related resonance strength. If the perturbation strength is time--independent $A_j(\theta) = A_j$, the above equations suggest the well-known resonance lines in rings, $k\nu_x+l\nu_y=p$ \cite{21}. Considering the fact that the perturbed space charge potential is self-consistently varying along with the beam oscillation from period to period, the phase advance of collective mode $\Phi_{j;k,l}$, derived from a particular perturbation pattern $A_j(s) x^{n-m}y^m$ in Eq.~(1),  is used to refer the oscillating characteristics of the space charge related time-dependent ``psedomultipoles". Thus, the explicit form of the perturbed potential pattern (Eq.~(\ref{eq4})) and resonance strength  (Eq.~(\ref{eq5}))  turn to  
\begin{eqnarray}\label{eq6}
V_n&=&\sum_{j=1}^n\sum_{k,l}A_j(\theta)\bar{V}_{k,l}(\theta)e^{i\Phi_{j;k,l}} \nonumber  \\
G_{k,l;p}&=&\frac{1}{2\pi}\int_0^L d\theta A_j(\theta)\bar{V}_{k,l}e^{-i((k\nu_x+l\nu_y-p)\theta-\Phi_{j;k,l})}. 
\end{eqnarray}
The resonance condition is modified as $k\nu_x+l\nu_y=p-\Phi_{j;k,l}/360^{\circ}$.  In linear accelerators with quasi-periodic structures, $p=1$.

\begin{figure}
    \centering
    \subfigure[]{
        \label{fig:subfig:4.2.1} 
        \includegraphics[width=1.5in]{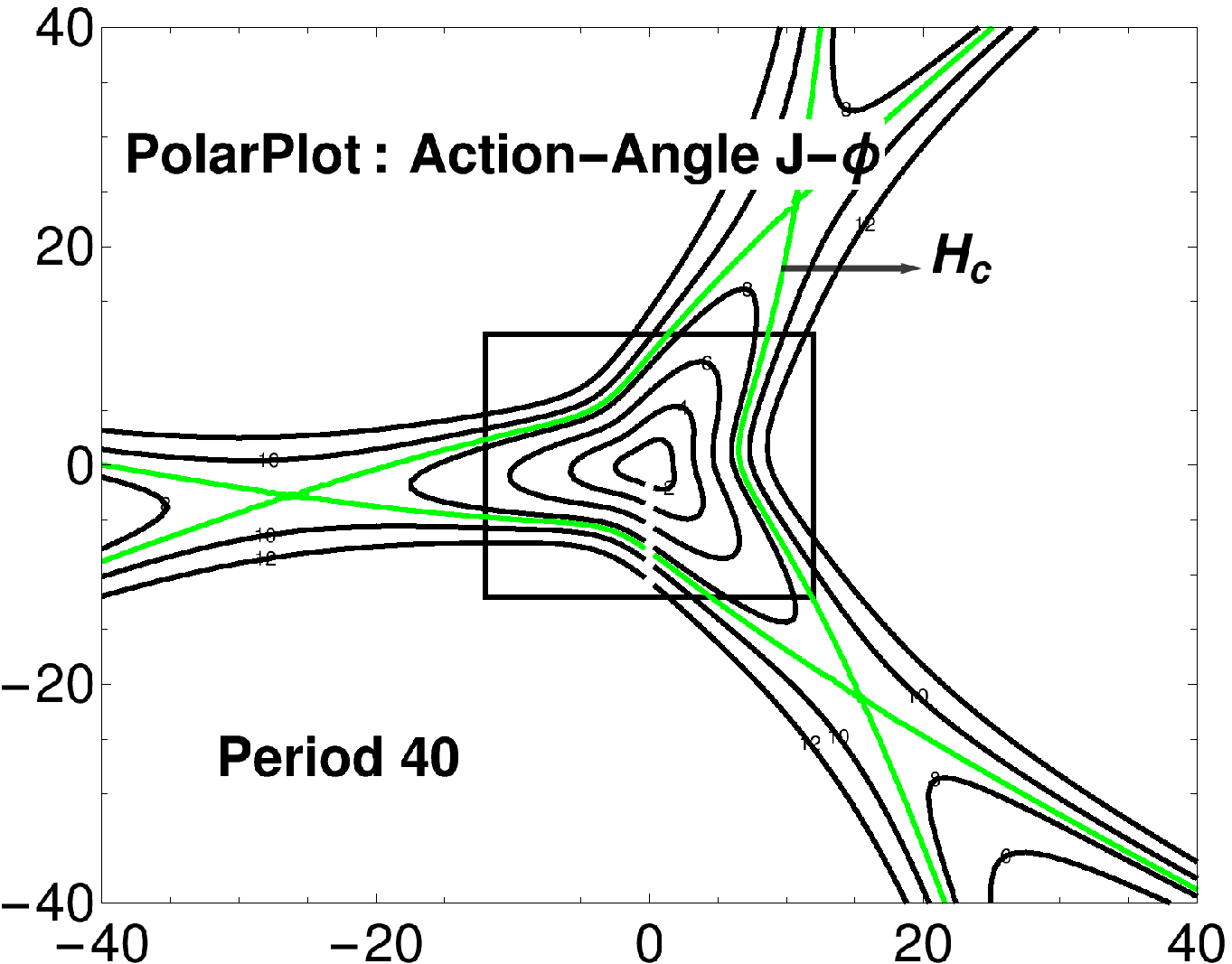}}
    \subfigure[]{
        \label{fig:subfig:4.2.2} 
        \includegraphics[width=1.5in]{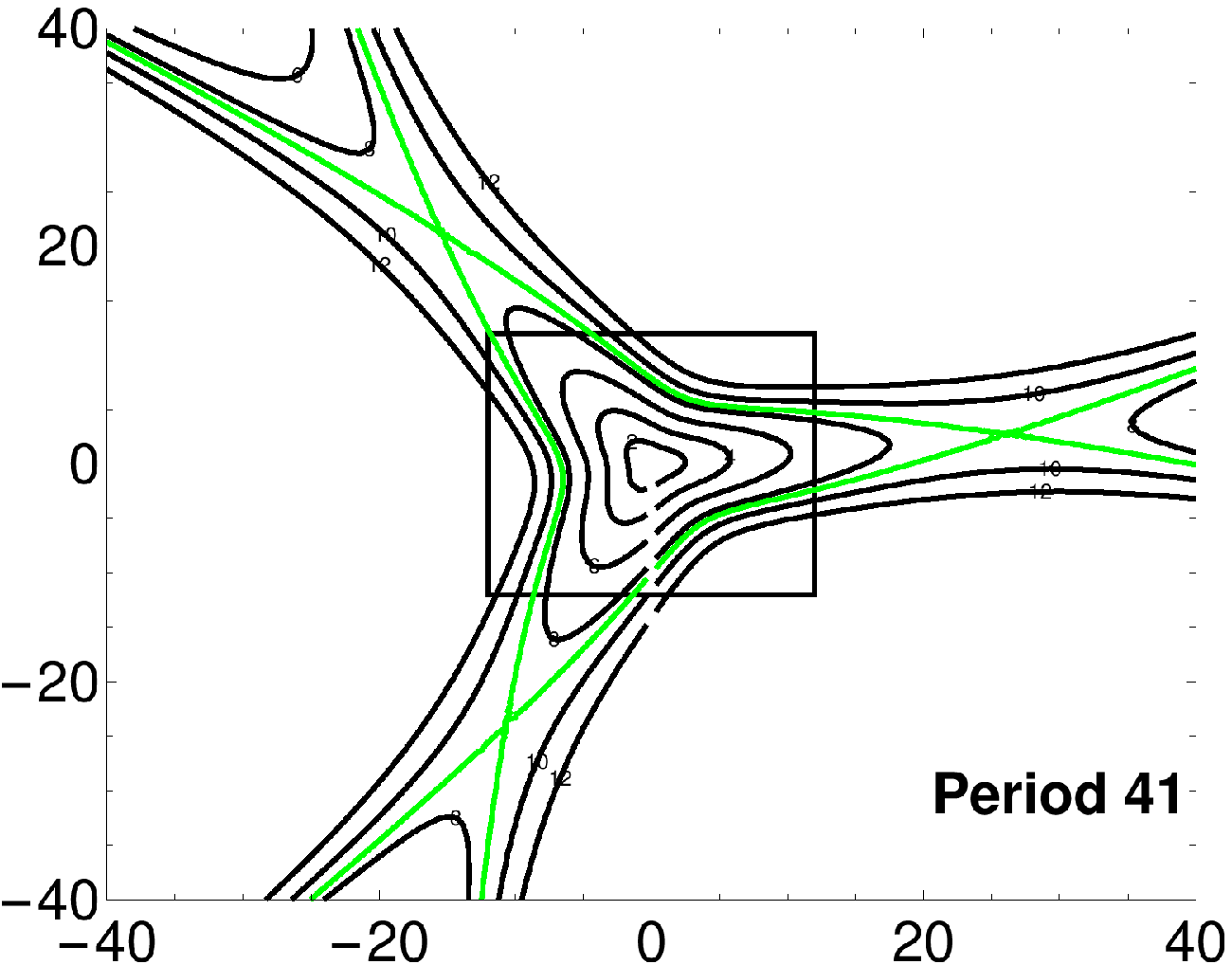}}
    \subfigure[]{
        \label{fig:subfig:4.2.3} 
        \includegraphics[width=1.5in]{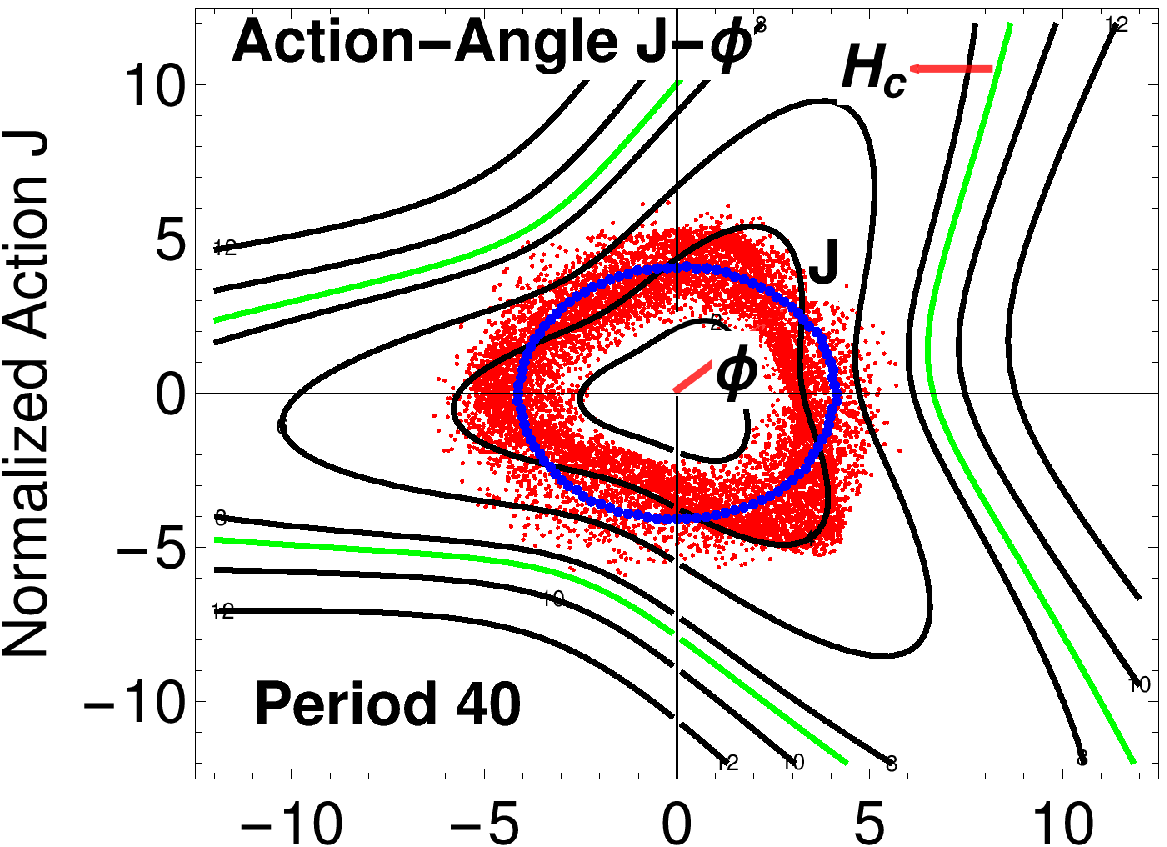}}
    \subfigure[]{
        \label{fig:subfig:4.2.4} 
        \includegraphics[width=1.5in]{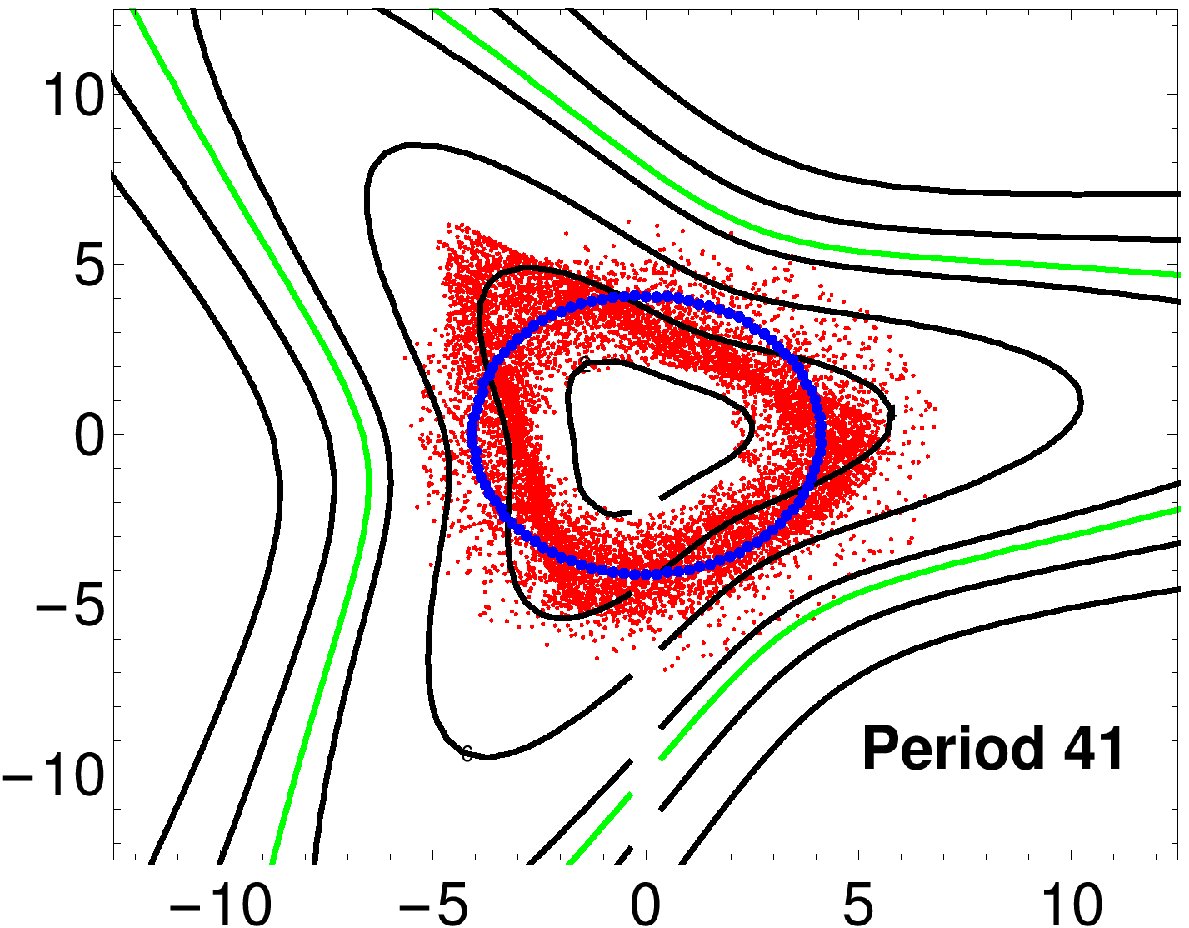}}
    \subfigure[]{
        \label{fig:subfig:4.2.5} 
        \includegraphics[width=1.5in]{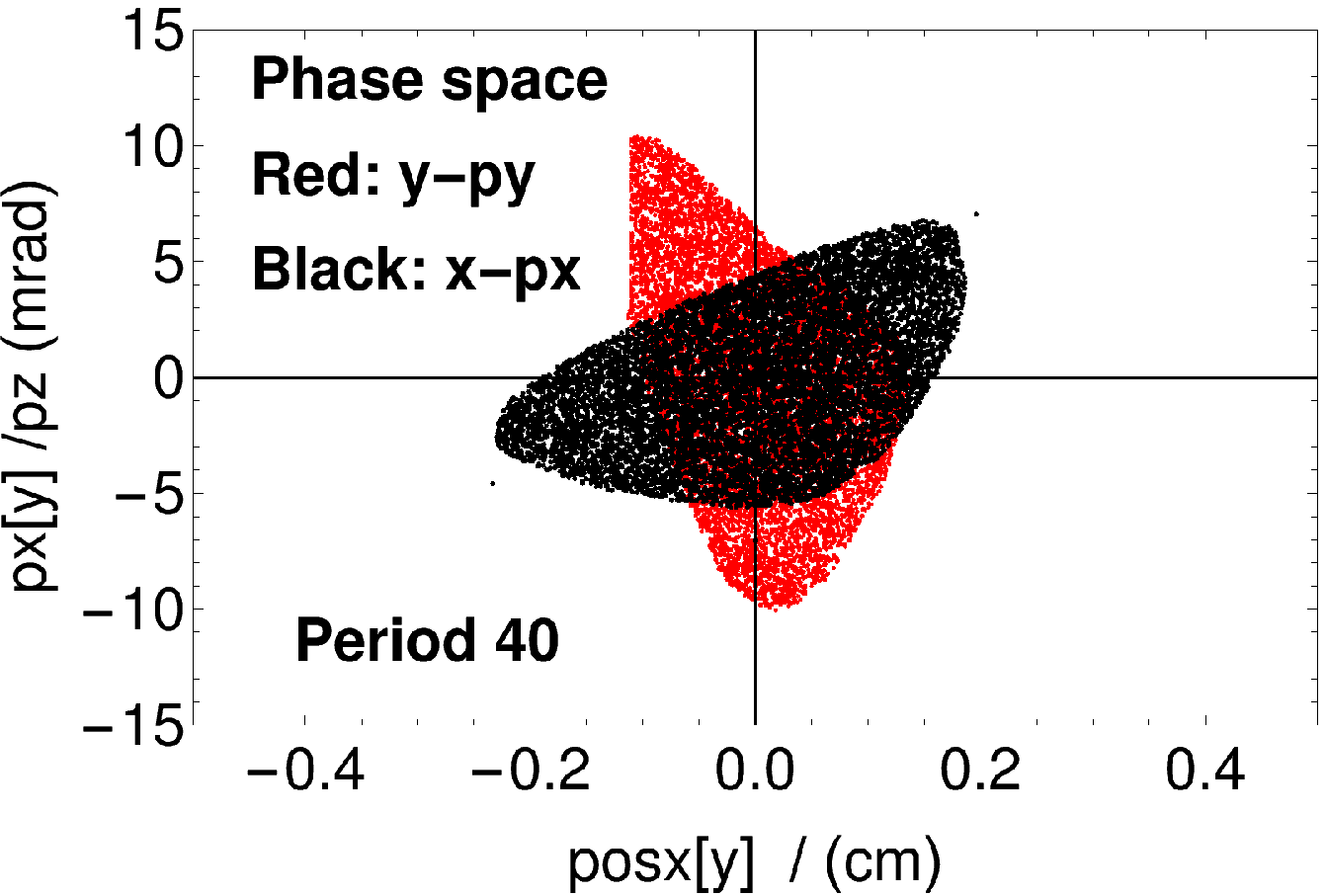}}
    \subfigure[]{
        \label{fig:subfig:4.2.6} 
        \includegraphics[width=1.5in]{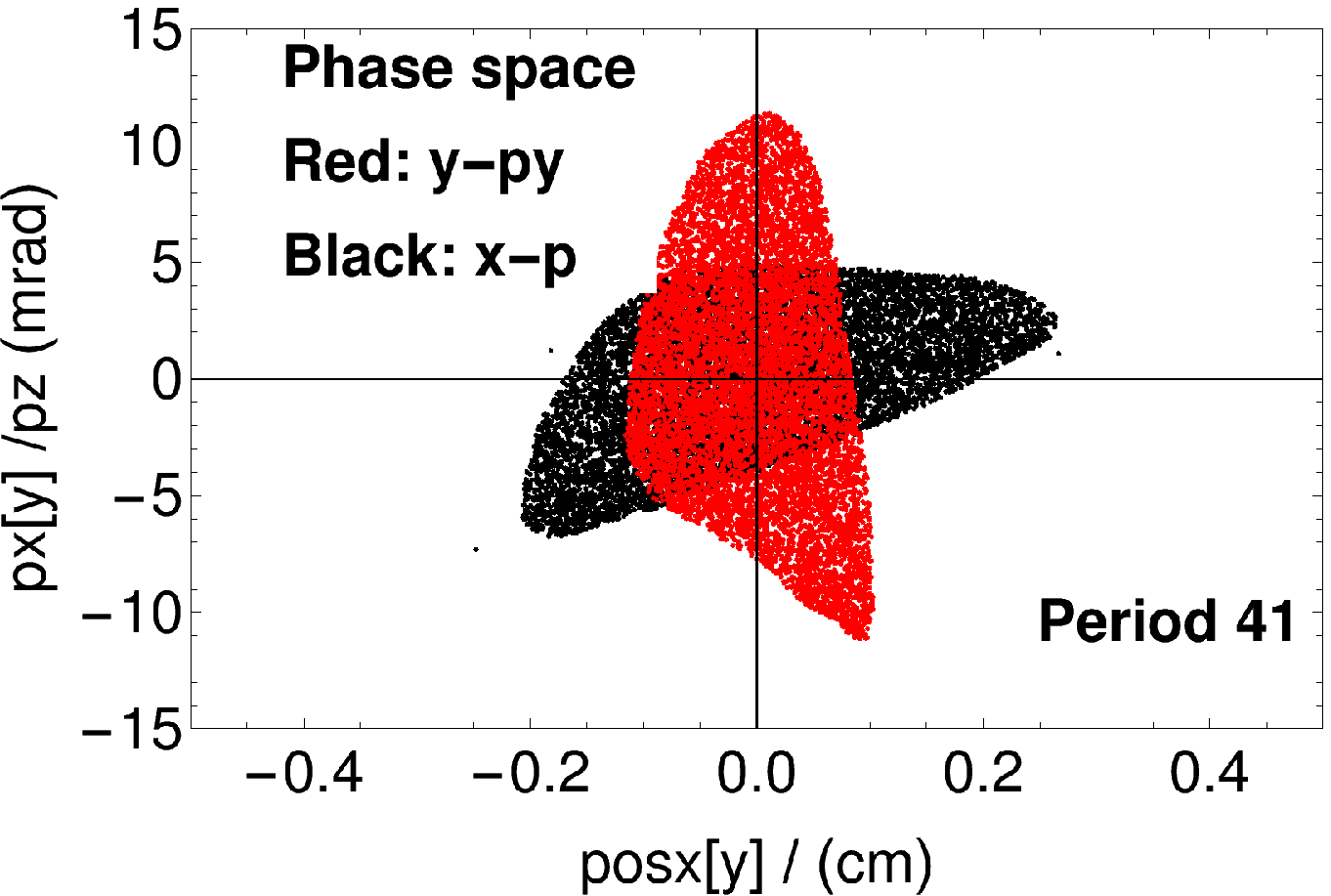}}
        \caption{\label{fig:4.2} The Hamiltonian tori of Eq.~(\ref{eq7})  with a perturbation strength of the form $G_{3,0;1}\sim 0.13J^{3/2}$ in polar action--angle phase space at period 40 (a) and 41 (b); beam distribution (red dots) and average action (blue dots) at action--angle phase space at    period 40 (c) and 41 (d); beam distribution in regular  ($x,p_x$), ($y,p_y$) phase space  at period 40 (e) and 41 (f). The orientation of the $(J,\phi)$ phase space  rotates at average rate of the transient betatron phase advance $\sigma$ per period; the  triangular  structure in regular phase space flips. The condition for PIC simulation in TOPO code is $\sigma_0=80^{\circ}$, $\sigma=50^{\circ}$ with KV initial beam in a FODO channel.} 
\end{figure}

To show the basic physical picture of the interaction between the single particle and collective modes, we chose a beam transport system with equal focusing in two degrees of freedom as a demonstration, in which 20K macro-particles with KV initial rms matched beam, where the normalized actions for all particles are initially equal  and located on a circle $J_0=\nu_x J_x+\nu_y J_y=4$,  are tracked by TOPO code \cite{8,24}. The initial beam condition is chosen in the 3rd order stop band $\sigma_0=80^{\circ}$, $\sigma=50^{\circ}$, as shown in Fig.~\ref{fig:subfig:4.1.2}, where it is the mode  $\Phi_{0;3,0}$ $(j=0,k=3,l=0)$, derived from potential with a pattern $\sim x^3$ and locked by the $180^{\circ}$ line \cite{23}, that supplies a sustained kick to a single particle as a resonance driving force \footnote{Similarly, interaction between particle and collective modes can be extended to arbitrary higher order.}. According to Eqs.~(\ref{eq3}, \ref{eq4}), the modified Hamiltonian is 
\begin{eqnarray}\label{eq7}
&&J(\phi_x,J_x,\phi_y,J_y;\theta)= \nonumber  \\
&& \nu_x J_x +\nu_y J_y+G_{3,0;p}\cos((3\nu_x-p)\theta+\Phi_{0;3,0}+\xi_{3,0;p}), \nonumber  \\
\end{eqnarray}
and resonance condition is $3\sigma_x=1-\Phi_{0;3,0}/360^{\circ}$ (or $3\sigma_x=\Phi_{0;3,0}$), which suggests particles with  phase advance  $\sigma_s\sim60^{\circ}$ can be captured by this resonance. Fig.~\ref{fig:4.2} shows  how the manifold of the invariant tori and the  beam distribution in phase space vary from period (left column: Period 40) to period (right column: Period 41) when this $3\sigma_x=\Phi_{0;3,0}$ resonance is substantially excited.  In polar coordinate system, Figs.~\ref{fig:subfig:4.2.1} and \ref{fig:subfig:4.2.2} show  the manifold of the Hamiltonian tori (black curves) and the the boundary separatrix $H_c(J,\phi)$ (green curve) near which particles can be easily kicked to a hyperbolic trajectory outside with large action.  Figs.~\ref{fig:subfig:4.2.3} and \ref{fig:subfig:4.2.4} clearly show that the particles (red dots) inside the separatrix ($J<H_c(J,\phi)$) closely follow the pattern of the bounded Hamiltonian tori in the action -- angle  ($J,\phi$) phase space. The blue dots  represent the average action $\bar{J}$ in each slice $\delta \phi$,  all at $\bar{J}\sim4$.  Figs.~\ref{fig:subfig:4.2.5} and \ref{fig:subfig:4.2.6} show the beam distribution in regular $(x,p_x)$ and $(y,p_y)$ phase space. Due to the self-consistence, the orientation of the Hamiltonian and beam distribution in  ($J,\phi$) phase space rotates at a average rate of the transient betatron phase advance $\sigma$ per period; as a consequence, the triangular  structure in regular phase space flips during one period as to $\Phi_{0;3,0}=180^{\circ}$. 

\if(false)
\begin{figure}
    \centering
    \subfigure[]{
        \label{fig:subfig:4.3.1} 
        \includegraphics[width=1.9in]{Fig4.3_2.eps}}
    \subfigure[]{
        \label{fig:subfig:4.3.2} 
        \includegraphics[width=1.1in]{Fig4.3_1.eps}}
        \caption{\label{fig:4.3}The beam distribution at period 18 with the parameter $\sigma=35^{\circ}$, $\sigma_0=80^{\circ}$ in action J and angle $\phi$ phase space. The orientation of the $J-\phi$ phase space  rotates at average rate of the transient betatron phase advance $\sigma$ per period.} 
\end{figure}
\fi

\begin{figure}
    \centering
    \subfigure[]{
        \label{fig:subfig:4.5.1} 
        \includegraphics[width=1.5in]{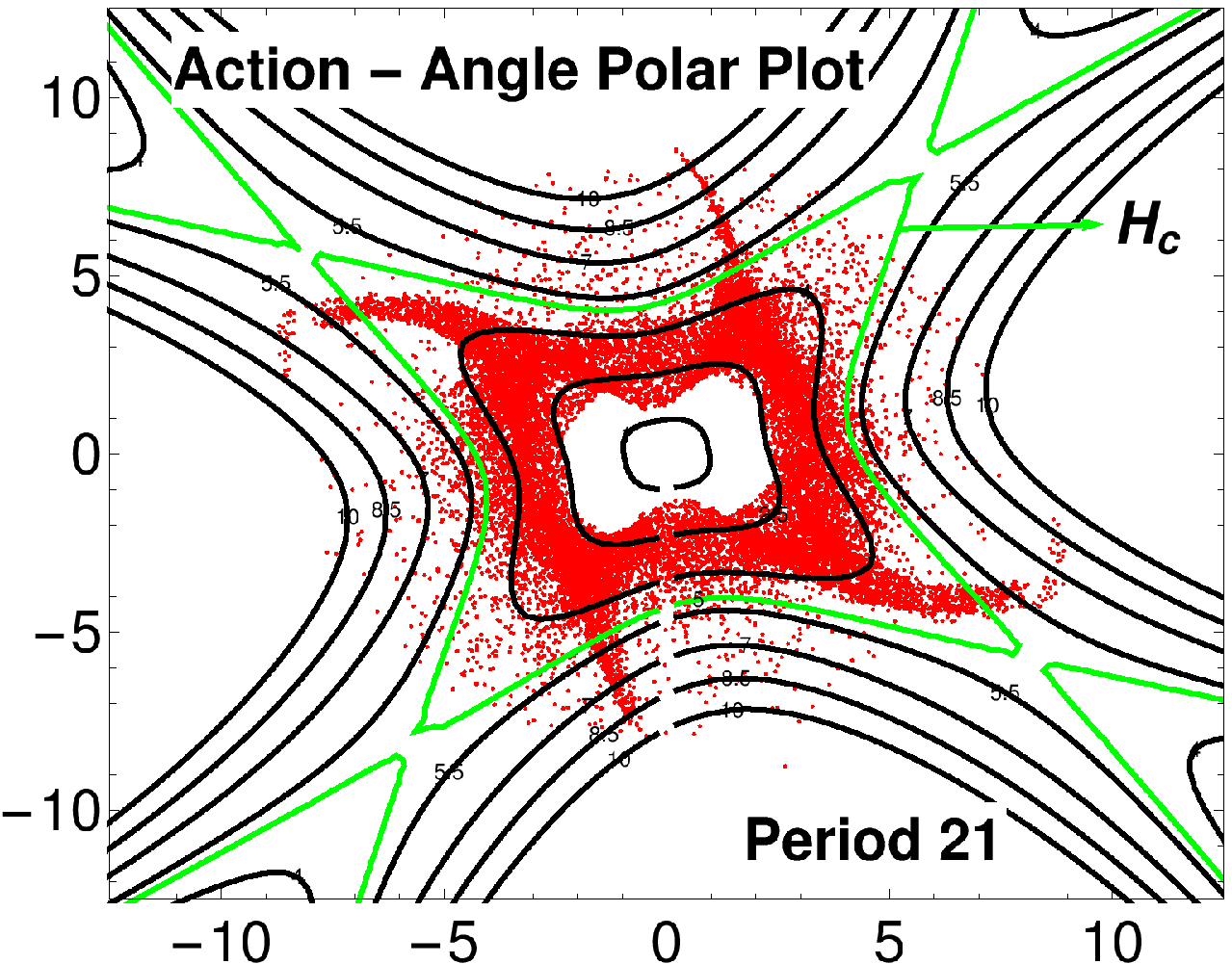}}
    \subfigure[]{
        \label{fig:subfig:4.5.2} 
        \includegraphics[width=1.5in]{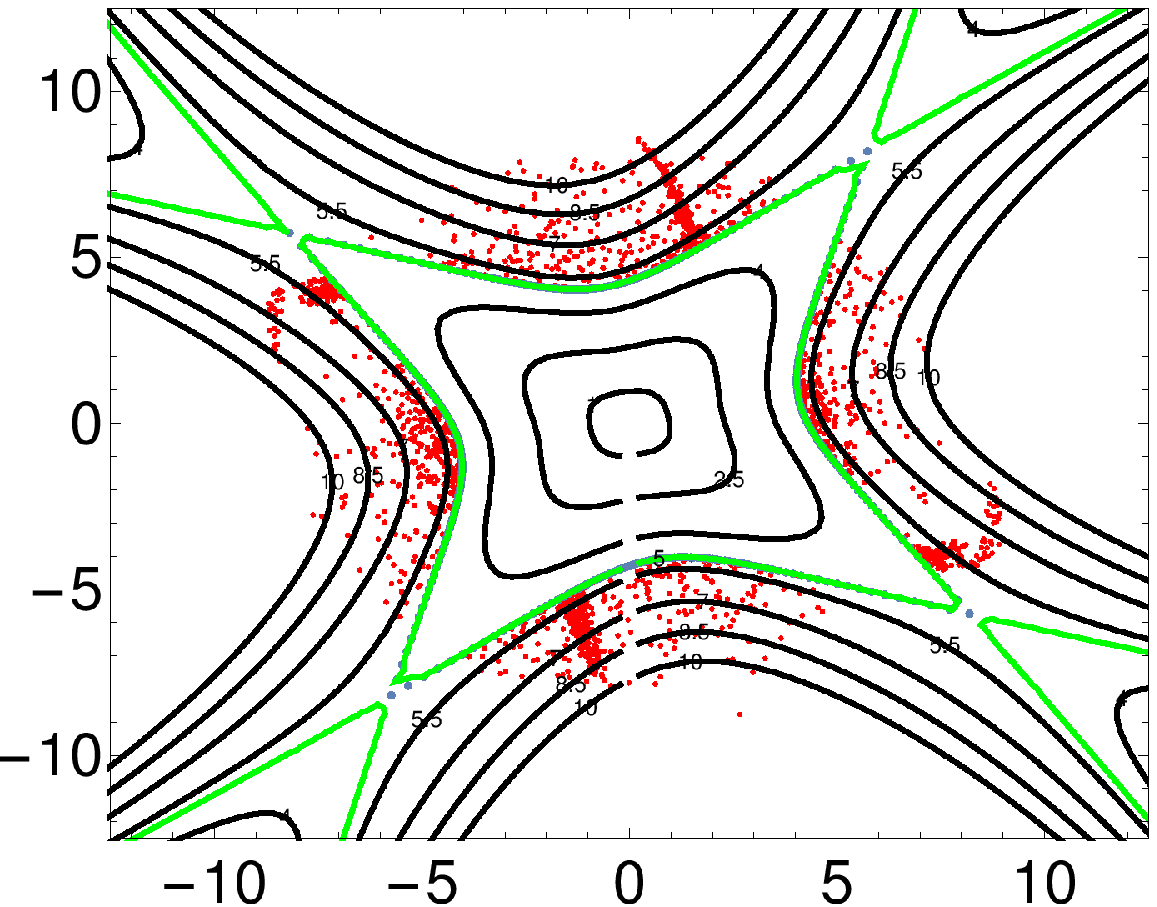}}
    \subfigure[]{
        \label{fig:subfig:4.5.3} 
        \includegraphics[width=1.5in]{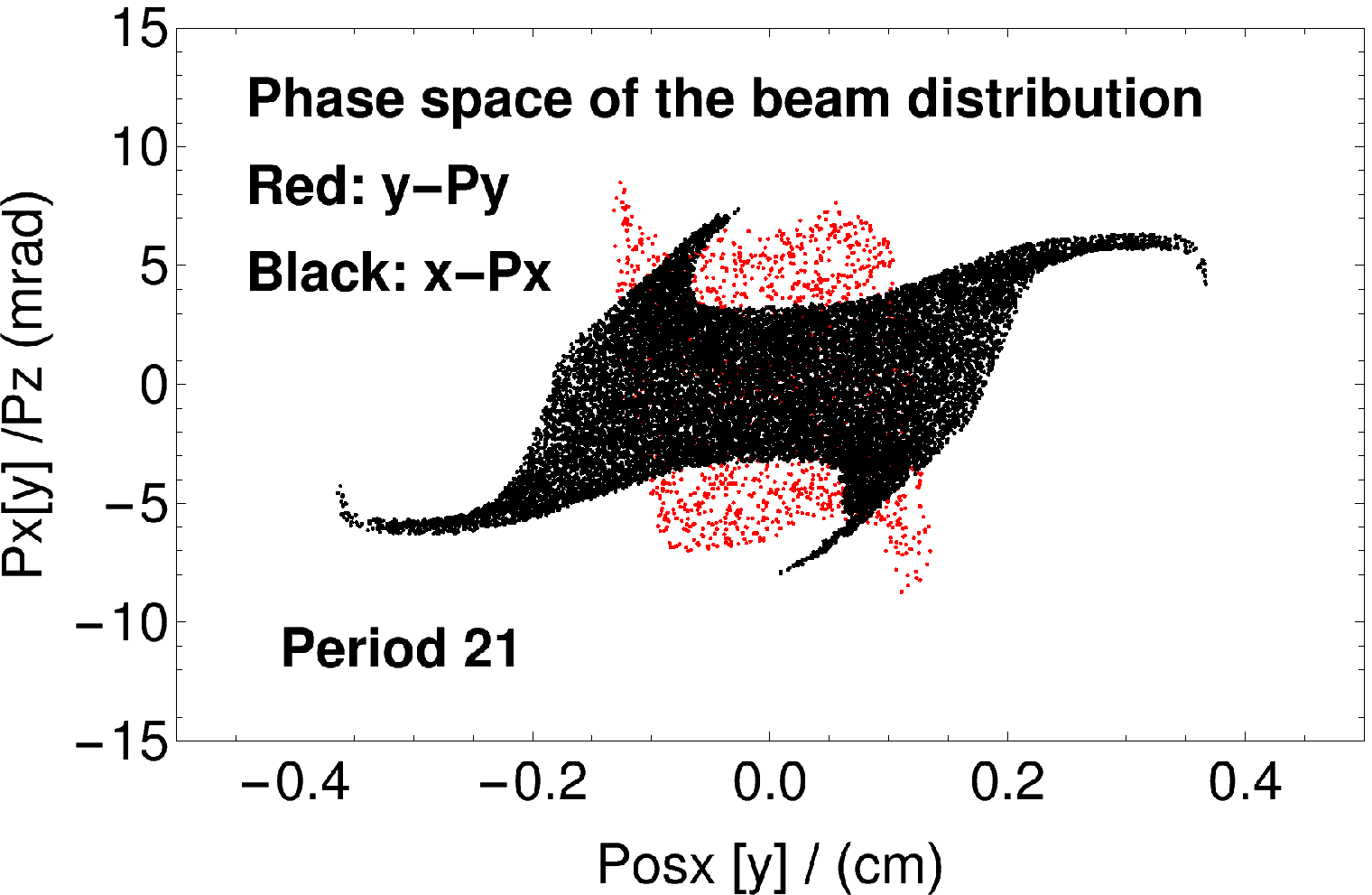}}
    \subfigure[]{
        \label{fig:subfig:4.5.4} 
        \includegraphics[width=1.5in]{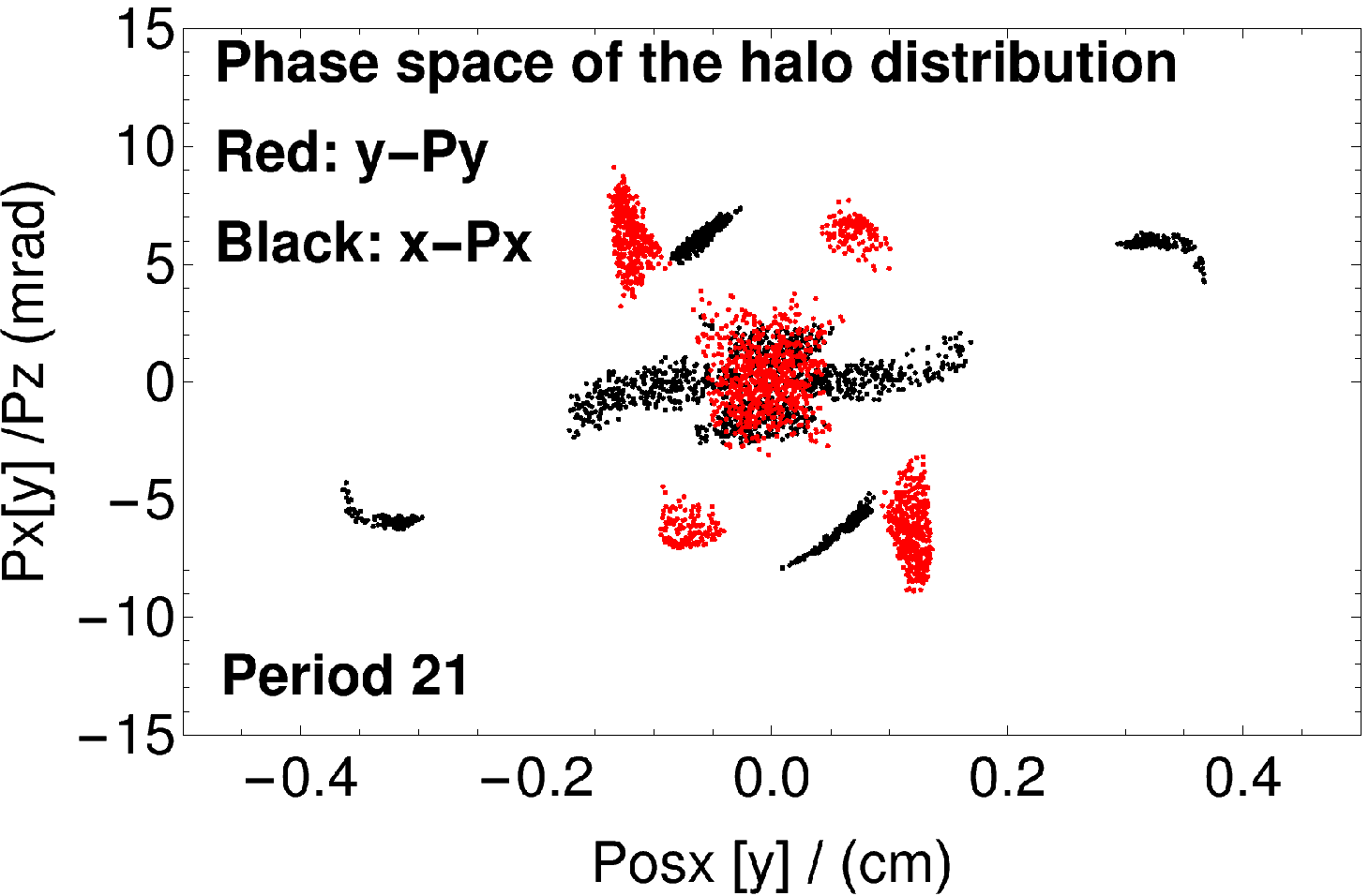}}
    \subfigure[]{
        \label{fig:subfig:4.5.5} 
        \includegraphics[width=1.5in]{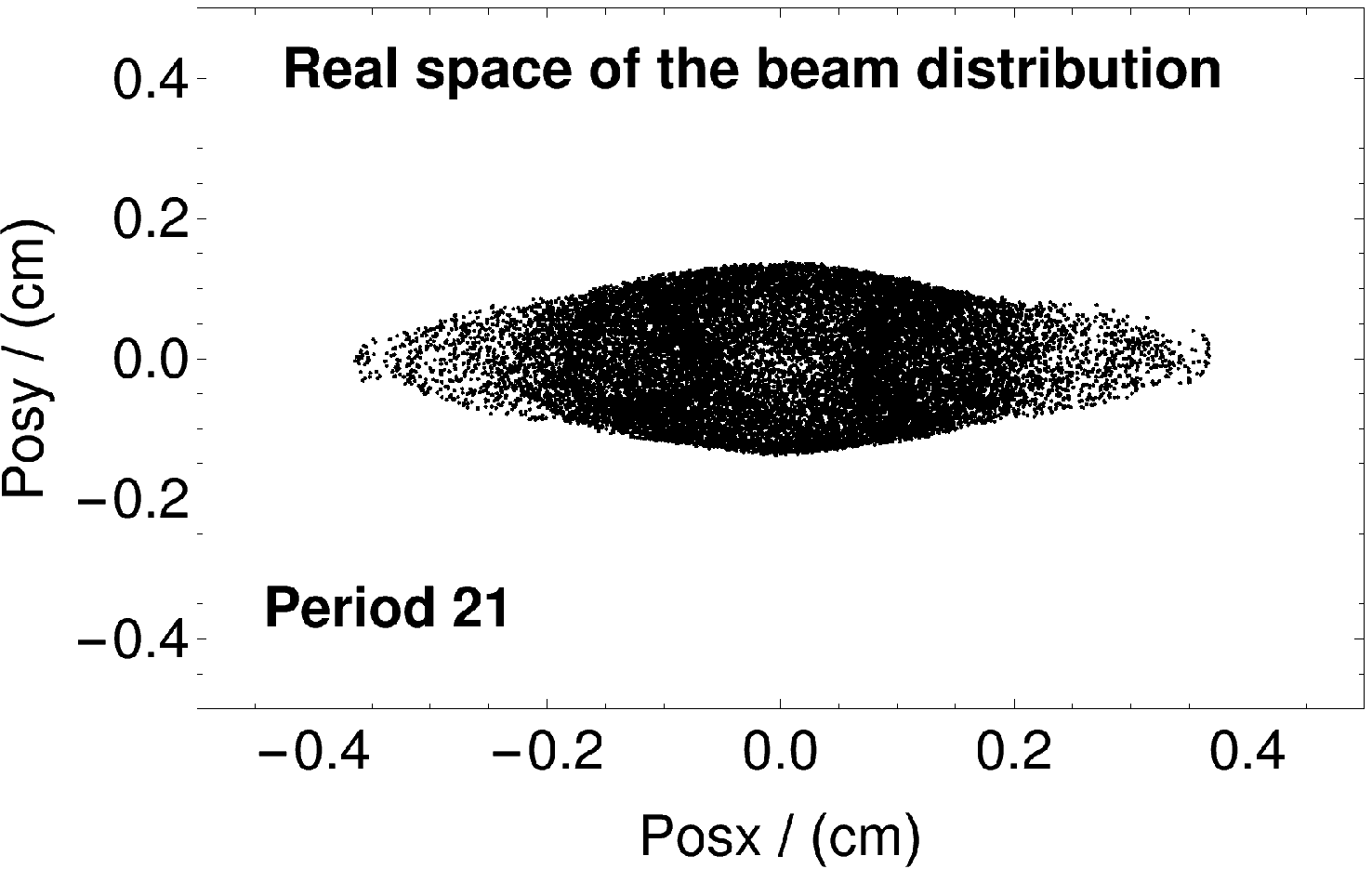}}
    \subfigure[]{
        \label{fig:subfig:4.5.6} 
        \includegraphics[width=1.5in]{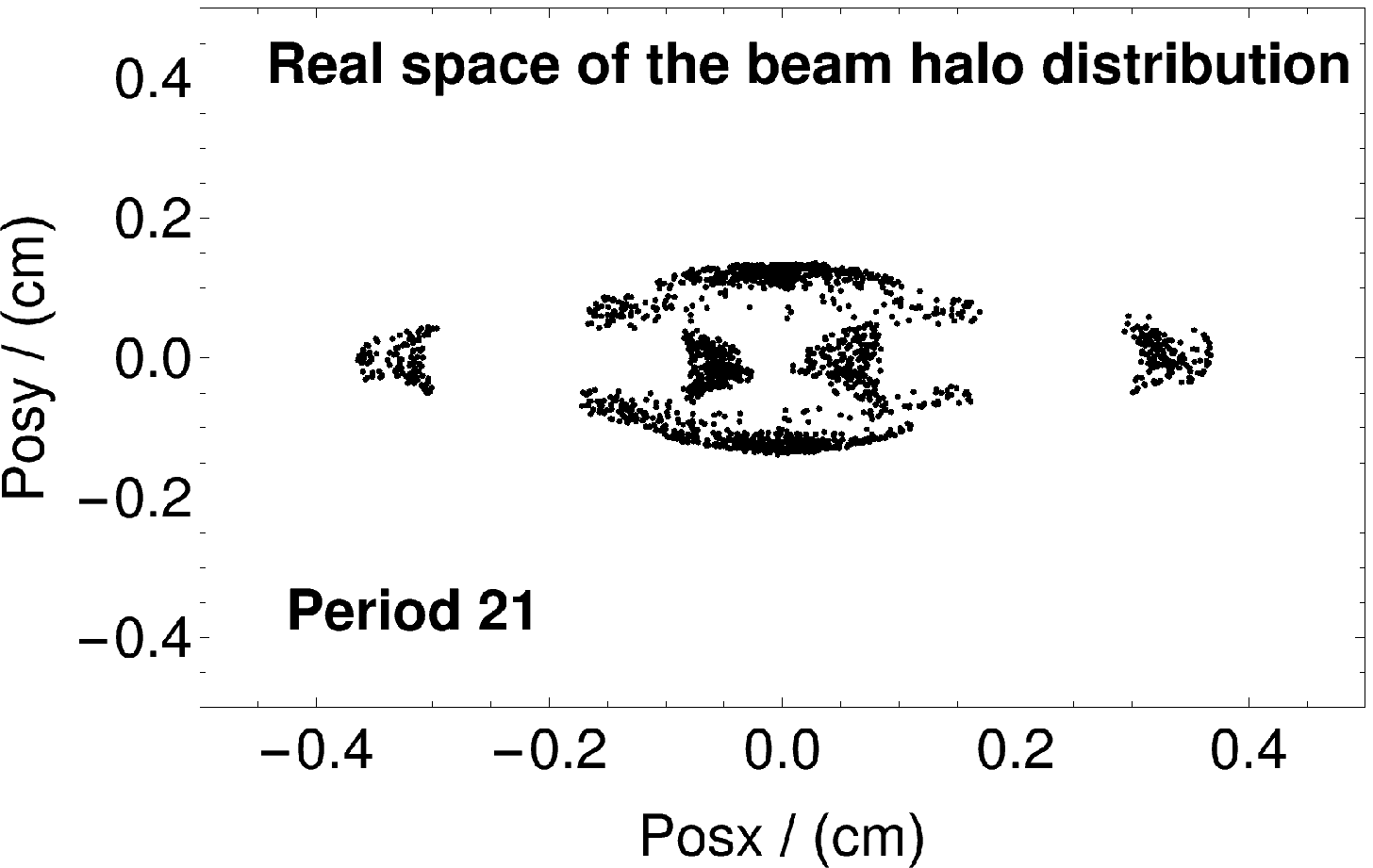}}
        \caption{\label{fig:4.5}  The left column represents beam profiles at the action-angle phase space (a), regular phase space (c) and real space (e) at period 21; the right column shows the beam profiles  including only the particles whose actions  beyond certain threshold $H_c(J,\phi)$ in action-angle phase space (b), regular phase space (d) and real space (f). The perturbation strength with a form $G_{4,0;1}\sim 0.05J^2$ in (a) and (b) is used to get the Hamiltonian tori. The condition for PIC simulation  in TOPO code is $\sigma_0=80^{\circ}$, $\sigma=35^{\circ}$ with KV initial beam in a FODO channel.  } 
\end{figure}

Usually, the ``halo" particles are described as the ``tail" particles in regular $(x,p_x)$ and $(y,p_y)$ phase space. In Figs.~\ref{fig:subfig:4.2.5} and ~\ref{fig:subfig:4.2.6}, the  ``tail" particles are commonly considered to be those located near the vertexes  of the triangular beam phase space. Here, we cite a new definition of beam halo: halo is particles outside the boundary $H_c(J,\phi)$, where they can acquire quite large action.  If desired, the definition might also include particles located in a range  near the the separatrix  $H_c(J,\phi)$  where they could  easily be kicked out from the stable tori inside the separatrix by a transient resonance effect to outside hyperbolic tori. To show the generality of this new halo description and the interaction between particle and collective modes discussed above,  another example with a  beam  $\sigma=35^{\circ}$, $\sigma_0=80^{\circ}$ injected into the 4th order structure resonance stop band (Fig.~\ref{fig:subfig:4.1.3}) is presented for discussion. There are two modes $\Phi_{0;4,0}$ and $\Phi_{2;2,2}$ locked in the $180^{\circ}$ line playing roles as sustained resonance driving force. For simplicity, we only discuss one mode $\Phi_{0;4,0}$ ($j=0, k=4, l=0$), and the resonance condition is $4\sigma_s\sim180^{\circ}$. 

The left column of Fig.~\ref{fig:4.5} shows the beam profile in action-angle phase space and the projected planes. The right column shows the related profile  including only the particle whose actions are  beyond the threshold $H_c(J,\phi)$. It shows that halo particles in   ($J,\phi$)  phase space (Fig.~\ref{fig:subfig:4.5.2}) include  all of the particles  located in the tail parts of the 4-fold structure in the projected 2D phase space (Fig.~\ref{fig:subfig:4.5.4}).  It indicates that the traditional geometrical  qualitative halo definition as ``particle outside of the core" in the projected 2D planes is only a partial  and subjective  description of  halo particles. On the projected phase space profile Fig.~\ref{fig:subfig:4.5.4}, halo particles in the inner region in  ($x,p_x$) phase space are located in the outer region in ($y,p_y$) phase space to meet the requirement that the total Hamiltonian is larger than $H_c(J,\phi)$. Particle distribution in real space $(x,y)$, Fig.~\ref{fig:subfig:4.5.6}, shows that collimators only can remove halo particles  which are at large positions in real space;  ``tail" particles will be regenerated  again since other particles are  hidden inside, Fig.~\ref{fig:subfig:4.5.4} and Fig.~\ref{fig:subfig:4.5.6},  which have small positions  but large momentum.

It is noteworthy that the approaches and understanding of the resonance between particle and collective mode due to space charge are very similar to the resonance study in rings due to lattice imperfection, except effects due to space charge have to be studied transiently to ensure the self-consistence.  It indicates that the space charge induced  perturbation strength $G_{k,l;p}$, Eq.~(\ref{eq6}), varies from period to period and the particle-collective mode  resonance condition becomes a dynamically varying quantity depending on the instantaneous beam distribution. More complicated particle motion can be foreseen especially when different orders of resonance are mixed, which make the analytical study almost impossible. Analytically it is not trivial to obtain the separatrix function $H_c(J,\phi)$. 

However, a practical analysis tool is needed to investigate the instantaneous state of the beam, and this is possible by simulation. Having the particle distribution in action-angle phase space, a table describing the normalized beam density as function of angle can be established, showing the beam density variation pattern along the angle. The perturbation strength $G_{k,l;p}$ in the Hamiltonian Eq.~(\ref{eq3}) is adjusted to agree with the table and then Eq.~(\ref{eq3}) is plotted to show $H_c$ and the contours as in Fig.~\ref{fig:4.2} and Fig.~\ref{fig:4.5}.

In summary, the above examples show the interaction between particle and collective modes, described in the action--angle phase space. The particles, pushed out to the  hyperbolic tori in action-angle phase space, can be safely ``defined" as beam halo with a distinct physical meaning, and includes all ``tail" particles visually estimated  in regular 2D phase space.  The effects of synchrotron motion (bunched beams) need to be studied separately, to investigate the expected mixing effect caused by the large  variation of phase advance $\sigma_s$ during the synchrotron oscillation.



Great thanks for discussion from Mei Bai. This work is supported by the Ministry of Science and Technology of China under Grant No.2014CB845501.

\bibliography{reference}
\end{document}